\documentclass[twocolumn]{aa}
\usepackage{graphicx}
\usepackage{txfonts}
\def\brfrac#1#2{\left(\dfrac{#1}{#2}\right)}
\def\dfrac#1#2{{\displaystyle\frac{\mathstrut #1}{#2}}}

\def\ea{{\rm et~al.\ }}

\def\Mdot{{\dot{M}}}

\def\Mbh{{M_{\rm BH}}}
\def\Msun{{M_\odot}}

\def\Ledd{{L_{\rm Edd}}}
\def\Leddc2{{L_{\rm Edd}/c^2}}

\def\Hb{{{\rm H} \beta}}
\begin{document}
   \title{Growth of massive black holes by super-Eddington accretion}

   \author{Toshihiro Kawaguchi\inst{1,}\inst{2} 
           \fnmsep\thanks{Present Address; 
  National Astronomical Observatory of Japan, 
  Tokyo 181-8588, Japan; 
  \email{kawaguti@optik.mtk.nao.ac.jp}},
          Kentaro Aoki\inst{3},
          Kouji Ohta\inst{4}
          \and
          Suzy Collin\inst{1}
          }

   \offprints{T. Kawaguchi}
   \authorrunning{T. Kawaguchi et al.}
   \titlerunning{BH growth in super-Eddington AGNs}

   \institute{LUTh/Observatoire de Paris-Meudon, 
              5 Place J. Janssen, 92195 Meudon, France 
%
%
             \and
             Postdoctoral Fellow of the Japan Society for the 
              Promotion of Science
             \and
	     Subaru Telescope, National Astronomical Observatory of Japan,
	     650 North A'ohoku Place, Hilo, HI 96729, USA 
%
%
             \and
	     Department of Astronomy, Graduate School of Science, 
	     Kyoto University, Sakyo-ku, Kyoto 606-8502, Japan 
%
%
             }

   \date{Received March 17, 2004; accepted April 30, 2004}

   \abstract{
{ 
Narrow-Line Seyfert 1 galaxies (NLS1s) and Narrow-Line quasars (NLQs)
seem to amount to $\sim 10-30 \%$ of active galactic
nuclei (AGNs) in the local universe.
Together with their average accretion rate,
we argue that a black hole (BH) growth by factor of $8-800$ 
happens in these super-Eddington accretion phase of AGNs.
Moreover, there is a possible, systematic underestimation of 
accretion rates (in the Eddington unit)
due to an overestimation 
of BH mass by massive accretion discs 
for super-Eddington objects.
If it is true, the factor of BH growth above 
may be larger by order(s) of magnitude.
In contrast, 
the growth factor expected in sub-Eddington phase
is only $\sim 2$.
Therefore, 
the cosmic BH growth by accretion is likely dominated 
by super-Eddington phase, rather than sub-Eddington phase 
which is the majority among AGNs.

This analysis is
based on the fraction and the average accretion rate 
of NLS1s and NLQs 
obtained for $z \lesssim 0.5$.
If those numbers are larger at higher redshift 
(where BHs were probably less grown), 
super-Eddington accretion would be even more important
in the context of cosmic BH growth history.
}
   }

   \maketitle
%

\section{Introduction}

Almost all galaxies in the local universe have 
supermassive black holes (BHs) in their nuclei,
not only active galactic nuclei (AGNs) but also  
apparently normal galaxies (Kormendy \& Richstone 1995; Magorrian \ea 1998).
Luminous quasars with massive BHs are observed till redshift
$z \approx 6$, when the universe was less than $10 \%$ of its recent age
(e.g. Fan \ea 2003; Vestergaard 2004).
However, it is not clear at all when those BHs have been formed and 
how their BH mass ($\Mbh$) have evolved.

Most of unobscured AGNs show broad 
Balmer lines of hydrogen 
(FWHM of $\Hb$ $>$ 2000 km s$^{-1}$),
and they are named as broad line Seyfert 1 galaxies (BLS1s) and quasars.
From several independent arguments,
the duration 
of an AGN is thought to be $\sim 10^8$ yr (Martini 2004 for a review).
And the average accretion rate $\Mdot$ of BLS1s/quasars is 
about $3 \, \Leddc2$ (e.g., Fig. 11 of Kawaguchi 2003),
where $\Ledd$ is the Eddington luminosity 
[$\sim 10^{46} (\Mbh / 10^8 \Msun) \, 
{\rm erg} \ {\rm s}^{-1}$]. 
Thus, BHs increase their mass during $\sim 10^8$ year only 
by a factor of $\sim 2$.

Then, what else can be responsible for the cosmic BH growth?
Semi-analytical approaches (e.g. Kauffmann \& Haehnelt 2000) employ
a scheme where a BH mass is increased not only by accretion
but also by merging of two massive BHs 
during a merger event of galaxies,
where the amount of BH masses is conserved.
However, the BH growth via merging is only expected
at high mass BHs (associated in most massive dark halos; Cattaneo 2002),
and thus this mechanism is not feasible for BH growth 
from a seed BH to a Seyfert-level BH ($\sim 10^{6-8} \Msun$).
Moreover,
discussions comparing the evolution of 
luminosity functions (LFs) with the volume
mass density of BHs 
(So\l tan 1982; Chokshi \& Turner 1992) 
infer that the BH growth by accretion process is enough to 
explain the local BH mass density 
(Fabian \& Iwasawa 1999; Yu \& Tremaine 2002;
Elvis, Risaliti \& Zamorani 2002).
Namely, 
merging of BHs might be
negligible for the cosmic BH growth (Marconi \ea 2004).

There is growing evidence that a certain 
class of AGNs has a higher 
Eddington ratio (ratio of bolometric to Eddington luminosity) 
among the AGN population;
Narrow Line Seyfert 1 galaxies (NLS1s) and their high luminosity 
analogue, 
Narrow-Line QSOs (NLQs).
They exhibit relatively narrow Balmer lines 
only slightly broader than the forbidden lines
(FWHM of $\Hb <$ 2000 km s$^{-1}$),
optical Fe II multiplet emission, 
and low $[O III]$--$H_\beta$ flux ratio, $<3$ (see Pogge 2000).
Strong soft X-ray excess
(e.g. Pounds, Done \& Osborne 1995; Boller, Brandt \& Fink 1996;
Boller \ea 2003),
and rapid X-ray variability (e.g. Otani \ea 1996; Leighly 1999; 
Boller \ea 2002; Gallo \ea 2004) are often observed. 
Those characteristics indicate that 
they have relatively small $\Mbh$ and larger $\Mdot / (\Leddc2)$ 
(e.g. Brandt \& Boller 1998; Hayashida 2000; Mineshige \ea 2000).

At super-Eddington accretion regime, 
the radiative efficiency per unit mass accretion 
is expected to decrease due to the onset of photon trapping 
(Begelman 1978).
As a result, 
the emergent luminosity 
from an accretion flow starts to saturate at around a few 
times $\Ledd$ 
(Abramowicz \ea 1988).
Thus, rapid BH growth is possible in super-Eddington sources, NLS1s/NLQs, 
(Rees 1992; Mineshige \ea 2000; Collin \ea 2002; Kawaguchi 2003).
However, 
the amount of BH growth
via super-Eddington accretion is quite ambiguous, 
since it strongly depends on the assumed duration of NLS1s/NLQs.
The terms, ``super-Eddington'' and ``sub-'', 
are used in this {\it Letter} when an accretion rate of an AGN 
is larger and smaller than $\sim 10 \Leddc2$, 
where the emergent bolometric
luminosity is about $(0.5-1)\, \Ledd$.

In this {\it Letter}, 
we assume that the ratio of the durations of a NLS1/NLQ 
and a BLS1/quasar may be proportional to the relative number ratio of 
these two classes.
Then, 
a simple algebra taking into account 
the fraction and average accretion rate of NLS1s/NLQs is 
discussed, in order to
evaluate qualitatively 
to what extent BHs can grow in super-Eddington accreting AGNs (\S 2).
It is followed by several discussions in the final section.


\section{Super-Eddington accretion and BH growth}

\subsection{Fraction and duration of NLS1 phase}

A fraction of AGNs are NLS1s and NLQs:
$\sim  11$\% objects among a heterogeneous sample (Marziani \ea 2003),
$\sim  15$\% among the Early Data Release of the 
Sloan Digital Sky Survey (SDSS; Williams \ea 2002),
and 
$\sim  (31-46)$\% among soft X-ray selected AGNs (Grupe \ea 1999; 
Grupe 2004; Salvato, Greiner \& Kuhlbrodt 2004).
The fraction can be even higher in Extreme UV band; 
8 NLS1s out of 14 radio-quiet AGNs (Edelson \ea 1999).
This wavelength dependency is likely due to the difference
of the spectral energy distribution between NLS1s/NLQs and BLS1s/quasars 
(Boller et al. 1996; 
Wang \ea 1996; Laor et al. 1997; see Fig. 14 of Kawaguchi 2003).
Most of those NLS1s have a redshift $z$ smaller than 0.5.
Here, we employ two numbers, $10 \%$ and $30 \%$, 
as the relative fraction of the NLS1/NLQ among AGNs.

In general, the duration (or the sum of multiple episodic phases) 
of an AGN is thought to be $\sim 10^8$ yr from
several independent arguments (e.g. Martini 2004;
Yu \& Tremaine 2002; Marconi \ea 2004).
The duration of each episodic phase must be longer than
$10^4$ yr, to explain the proximity effect (e.g. Bajtlik, 
Duncan \& Ostriker 1988) and the sizes of ionization-bounded
narrow-line regions (Bennert \ea 2002).

If statistical properties of AGNs (e.g. relative number ratio of 
super-Eddington to sub-Eddington sources) are quasi-steady, 
the fraction of NLS1s inferred above
will be proportional to the duration of NLS1/NLQ phase.
In fact, the evolution of AGNs is not so rapid in this
redshift range, $z = 0.5 \rightarrow 0$, 
taking $5 \,$Gyr:
the number density decreases by a factor of 3, 
and the knee of LFs stays at almost the same luminosity 
(e.g. Boyle \ea 2000; Miyaji \ea 2001; Ueda \ea 2003).

Then, let us simply assume that each NLS1 (or NLQ) 
has a duration of $\sim 10^7$yr or $\sim 3 \times 10^7$yr.
This number ($\sim 10^7$yr) was briefly discussed 
together with the fraction of NLS1s/NLQs 
also in a thesis (Grupe 1996).
A characteristic timescale for BH growth can be roughly 
evaluated by (Salpeter 1964) 
\begin{equation}
 \label{eq_grow_intro}
  \Mbh / \Mdot
  = 4.5 \times 10^8 \brfrac{\Mdot}{\Leddc2}^{-1} {\rm year}.
\end{equation}
If BH growth mainly occures in NLS1/NLQ phase,
the timescale of the order of $10^7$yr implies that 
 $\Mdot \gtrsim 45 \, \Leddc2$. 

\subsection{Accretion rate of NLS1s}

The BH mass and gas accretion rate of AGNs onto their central BHs can be 
estimated using 
the observed relation between optical luminosity
and the size of broad line region (Wandel, Peterson \& Malkan 1999;
Kaspi \ea 2000).
Comparing 
optical luminosity of NLS1s and NLQs between
observation and theoretical models, 
along with BH masses inferred from 
$H \beta$ and [O III] widths, 
their accretion rate is 
about $100 \Leddc2$, on average 
(Fig.\ 11 in Kawaguchi 2003; Collin \ea 2002).

Alternative and independent method to evaluate $\Mbh$
of NLS1s/NLQs is spectral fitting of slim disk models 
at soft X-ray range (Wang \& Netzer 2003).
Moreover, if one tries to fit the broad-band spectra at optical--X-ray band,
not only $\Mbh$ but also $\Mdot$ are estimated 
simultaneously (Kawaguchi 2003; Kawaguchi, Pierens \& Hur\'{e} 2004;
Kawaguchi, Matsumoto \& Leighly in prep.; see Kawaguchi 2004).
Currently, these models are the only ones
that explain broadband SEDs of NLS1s, in our best knowledge.
This kind of estimations indicate $\Mbh$ and $\Mdot$ which are similar to
those derived by the line widths in the previous paragraph.
Spectral fitting is relatively time consuming, but this success implies
that we can use the line widths for quick estimations, as the first
order.

\subsection{Linear growth of BHs}

Now, we start with an assumption that the gas accretion rate onto
a central BH is almost constant: $\Mdot \approx$ const.
For instance, a following scenario is possible for 
the constant-$\Mdot$ case:
large amount of gas accretion (super-Eddington accretion phase $=$ 
NLS1/NLQ-phase) may start 
induced by an unknown trigger.
With a growing BH mass, the accretion rate in unit of the Eddington 
rate will be reduced down to sub-Eddington phase 
(broad line Seyferts and quasars).
Accretion rate (in Eddington unit) will decrease continuously,
and eventually it will become lower than $\sim 0.2 \Leddc2$, 
followed by a transition from a radiatively efficient to an inefficient flow
(i.e. optically-thin advective flow; 
e.g. Meyer, Liu \& Meyer-Hofmeister 2000), such as Low Luminosity AGNs.

Employing this view, the growth of $\Mbh$ with time $t$ 
is linear as is expressed below:
\begin{equation}
 \label{eq_grow_lin}
  \Mbh (t)
  = \Mbh (t=0) + \Mdot \times t.
\end{equation}
Substituting the average accretion rate of NLS1s/NLQs
to $\Mdot$, we have 
\begin{equation}
 \label{eq_grow_lin_res}
  \Mbh (t=10^7 {\rm yr})
  \approx 8 \times \Mbh (t=0),\ {\rm and}
\end{equation}
\vspace*{-4mm}
\begin{eqnarray}
 \label{eq_grow_lin_res2}
  \Mbh (t=3 \times 10^7 {\rm yr})
  &\approx& 23 \times \Mbh (t=0),\\
  & & \hspace*{4mm} {\rm if} \ \Mdot = 100 \, 
   \frac{\Ledd (t=10^{6.5} {\rm yr})}{c^2}. \nonumber
\end{eqnarray}

On the other hand, in a sub-Eddington phase
the factor of BH growth during $\sim 10^8$ yr is not so significant:
\begin{eqnarray}
  \Mbh (t=10^8 {\rm yr})
  &=& 1.8 \times \Mbh (t=0), \nonumber \\
  & & \hspace*{10mm} {\rm if} \ \Mdot = 3 \, 
   \frac{\Ledd (t=10^{7.5} {\rm yr})}{c^2}. \nonumber
\end{eqnarray}

\subsection{Exponential growth of BHs}

Next, another assumption on accretion rate is that
$\Mdot / (\Leddc2)$ may be constant with time by 
some unknown self-regulation mechanism(s).
In this case, sub-Eddington phase does not follow at
the end of super-Eddington phase, contrary to the previous case.
Instead, in each trigger of AGN activity, $\Mdot / (\Leddc2)$
is (somehow randomly) chosen: if a nuclei has a super-Eddington
(sub-) accretion rate it will appear as a NLS1/NLQ 
(a broad-line Seyfert/quasar).
The growth of $\Mbh$ will be exponential with this assumption,
and is written as,
\begin{eqnarray}
 \label{eq_grow_exp}
  \Mbh (t)
  &=& \Mbh (t=0) \times \exp \left[t \times \brfrac{\Mdot}{\Mbh} 
			     \right] \nonumber\\
  &=& \Mbh (t=0) \times \exp \left[\brfrac{t}{4.5 \times 10^8 {\rm yr}} 
			    \brfrac{\Mdot}{\Leddc2} \right].
\end{eqnarray}
Substituting the average accretion rate (\S 2.2), we get
\begin{equation}
 \label{eq_grow_exp_res}
  \Mbh (t=10^7 {\rm yr})
  \approx 9 \times \Mbh (t=0),\ {\rm and}
\end{equation}
\vspace*{-4mm}
\begin{equation}
 \label{eq_grow_exp_res2}
  \Mbh (t=3 \times 10^7 {\rm yr})
  \approx 800 \times \Mbh (t=0).
\end{equation}

Given the average accretion rate  ($\sim 3 \Leddc2$)
and duration ($\sim 10^8$yr) 
of BLS1s/quasars,
a BH can grow by a factor of 1.9 in the sub-Eddington phase.

\section{Discussions and conclusions}

In principle, a BH mass based on line widths is the mass
inside the broad line region, and does not necessarily equal to
the true $\Mbh$.
Outer part of accretion disks 
in NLS1s/NLQs, where optical continuum is emitted,  
are self-gravitating (Kawaguchi \ea 2004).
Such disks are quite massive, 
and $\Mbh$ based on line widths can be systematically
overestimated by a factor of $\sim 2$. 
If this is taken account into $\Mbh$ and $\Mdot$ estimations,
the average accretion rate of NLS1s/NLQs might be
$ 200 \Leddc2$, rather than $100 \Leddc2$.
Thus, 
the factor of BH growth presented in 
\S 2.4 (eqs.~\ref{eq_grow_exp_res}, \ref{eq_grow_exp_res2}) 
will be multiplied by themselves.
While, it is too efficient to grow a BH 
with this accretion rate for a constant-$\Mdot$ case (\S 2.3).
For sub-Eddington systems, on the other hand, there is no 
constraints upon the presence of outer, massive, self-gravitating 
disks, since such region (if exists) radiates at Near-IR where
the emission from dusty torus dominates.

We have utilized 
the relative fraction and 
the average accretion rate 
of NLS1s/NLQs obtained for $z \lesssim 0.5$.
Apparently, the Eddington ratios
of AGNs detected in SDSS tend to increase with redshift 
(McLure \& Dunlop 2003).
Thus, 
BH growth via super-Eddington accretion would be much more important
in the young universe than discussed above.

The tight correlation between spheroid mass of galaxies and 
BH mass indicates that they co-evolve.
Indeed, high star formation activities in host galaxies of NLS1s/NLQs are
inferred (e.g. Moran, Halpern \& Helfand 1996; Kawaguchi \& Aoki 2001).
Since the gas accretion rate of NLS1s/NLQs is $\sim 2 \Msun \, {\rm yr}^{-1}$ 
at most (Collin \& Kawaguchi 2004), 
a small fraction of 
mass loss
by starburst streaming toward the central BH
is enough to form a NLS1/NLQ.

We note that the super-Eddington accretion phase (with $L \lesssim \Ledd$)
discussed here is not a surprising assumption in the context of
cosmic structure formation history.
In semi-analytical approaches (e.g. Kauffmann \& Haehnelt 2000),
 for instance, most of accretion events happen in super-Eddington
regime if the episodic phase of an AGN is assumed to be less than $10^8$yr 
(Cattaneo 2001).

For quasars, with their bolometric 
luminosity $L \gtrsim 10^{46}$ erg/s, 
LFs can be obtained by direct observations to high 
redshifts in a wide range of wavebands.
Comparing the integration of those LFs with the volume
mass density of BHs in the local universe, 
the radiative efficiency 
seems to be quite high ($\gtrsim 0.15$)
in those luminous objects 
(e.g. Yu \& Tremaine 2002;
Elvis, Risaliti \& Zamorani 2002).
Such a high efficiency indicates 
that most BHs of quasars 
are rotating with sub-Eddington accretion rates. 
However, the mass density of local BHs are dominated by 
quite massive BHs at 
the knee of mass functions or LFs.
Therefore, constraints upon BH growth based on above discussions are likely 
valid 
at high mass end; e.g. BH growth from $10^8 \Msun$ to 
$10^{10} \Msun$.

On the other hand,
BH growth for lower luminous objects with 
$\Mbh \lesssim 10^8 \Msun$ 
is quite unclear. 
The way BHs grow in Seyferts
is not necesarilly the same as that of quasars.
Actually,
Eddington ratios of 
AGNs seems to be higher for objects with smaller $\Mbh$ 
(e.g. Collin \ea 2002; Collin \& Kawaguchi 2004).
Yu \& Tremaine (2002) argued 
that relatively low luminous AGNs
should have less radiative efficiency, $\lesssim 0.1$.
Thus, super-Eddington accretion (which is expected to have
very low efficiency) 
seems to play an important role in 
BH growth from seed BHs 
to $\sim 10^8 \Msun$. 
This is not in conflict with the idea above that BHs in quasars
grow with high radiative efficiency (via sub-Eddington accretion).

Given that a BH mass increases 
by the factors mentioned above, 
is it enough
to grow BHs found in quasars with $\sim 10^{8-10} \Msun$ 
from seed BHs?
It may be possible to make such supermassive BHs 
if i) the mass of seed BHs at the center of galaxies is large enough 
(e.g. $\gtrsim 10^6 \Msun$), and/or 
ii) the duration and average accretion rate of super-Eddinton
accreting AGNs are much larger at high redshift than those estimated for local 
objects.
Instead, iii) BH growth at the highest mass end may be indeed 
dominated by merging events as expected in semi-analytical
approaches (Cattaneo 2002).

In summary, 
we have argued that the growth of black hole (BH) mass by factor 
of $8-800$ happens in the super-Eddington accretion phase.
There is a possible, systematic underestimation of $\Mdot / (\Leddc2)$ 
by a factor of $\sim 2$ due to an overestimation 
of BH mass by massive accretion discs 
for super-Eddington objects.
If it is true, the factor of BH growth above 
may be larger by order(s) of magnitude.
On the other hand,
the growth factor expected in
sub-Eddington active galactic nuclei (AGNs) is only $\sim 2$.
Therefore, 
the cosmic BH growth by accretion is dominated by super-Eddington phase,
rather than sub-Eddington phase which is the majority among AGNs.
These values are based on the fraction and the average accretion rate 
of Narrow-Line Seyfert 1 galaxies and Narrow-Line quasars 
obtained for $z \lesssim 0.5$.
If those numbers are larger at higher redshift (where BHs are less grown), 
BH growth via super-Eddington accretion would be even more important.

For further analysis, we need i)
volume limited samples of AGNs to derive an accurate relative
fraction of NLS1s/NLQs, and ii) 
constraints on $z-$dependency of the relative fraction and
the average $\Mdot/(\Leddc2)$ to determine the importance
of super-Eddington accretion in BH growth history.
In this {\it Letter}, 
we divide radio-quiet type I AGNs into two subclasses:
objects with $\Mdot \gtrsim$ or $\lesssim 10 \Leddc2$.
It would be more appropriate to divide into fine bins
for dealing with the observed continuous distribution of 
$\Mdot/(\Leddc2)$;
e.g. bins for $10-100 \Leddc2$ and $100-1000 \Leddc2$, etc.
For this aim, iii) more sophysticated (with less uncertainty) methods
for $\Mdot/(\Leddc2)$ estimations are necesary.

\begin{acknowledgements}

We thank Masayuki Umemura, Nozomu Kawakatu 
and Andrea Cattaneo 
for illuminating discussions. 
Useful comments by the referee, Thomas Boller, are also
appreciated.
T.K.\ is supported by the Japan
Society for the Promotion of Science (JSPS) 
Postdoctoral Fellowships for Research Abroad (464).

\end{acknowledgements}

\end{document}